\documentclass{article}

\PassOptionsToPackage{numbers, compress}{natbib}
\usepackage[final]{neurips_2020}
\usepackage[utf8]{inputenc} % allow utf-8 input
\usepackage[T1]{fontenc}    % use 8-bit T1 fonts
\usepackage{hyperref}       % hyperlinks
\usepackage{url}            % simple URL typesetting
\usepackage{booktabs}       % professional-quality tables
\usepackage{amsfonts}       % blackboard math symbols
\usepackage{nicefrac}       % compact symbols for 1/2, etc.
\usepackage{microtype}      % microtypography
\usepackage{algorithm}
\usepackage{algorithmic}
\usepackage{mathtools}

\usepackage{bm}

\usepackage{graphicx}
\usepackage{float}
\usepackage{xcolor}

\title{FinRL: A Deep Reinforcement Learning Library for Automated Stock Trading in Quantitative Finance}

\author{Xiao-Yang Liu$^{1*}$,
       Hongyang Yang$^{2,3}$\thanks{Equal contribution.},~
       Qian Chen$^{4,2}$,
       Runjia Zhang$^3$,\\
       \textbf{Liuqing Yang$^3$,
       Bowen Xiao$^5$,
       Christina Dan Wang$^6$\thanks{Christina Dan Wang is supported in part by National Natural Science Foundation of China (NNSFC) grant 11901395 and Shanghai Pujiang Program, China 19PJ1408200.},~
       }\\
       $^1$Electrical Engineering, $^2$Department of Statistics, $^3$Computer Science, Columbia University,\\
       $^3$AI4Finance LLC., USA, $^4$Ion Media Networks, USA,\\
       $^5$Department of Computing, Imperial College,
       $^6$New York University (Shanghai)\\
       Emails: \{XL2427, HY2500, QC2231,
       LY2335\}@columbia.edu, \\
       info@ai4finance.net, bowen.xiao20@imperial.ac.uk, christina.wang@nyu.edu \\     
       }

\begin{document}

\maketitle

\begin{abstract}

  As deep reinforcement learning (DRL) has been recognized as an effective approach in quantitative finance, getting hands-on experiences is attractive to beginners. However, to train a practical DRL trading agent that \textit{decides where to trade, at what price, and what quantity} involves error-prone and arduous development and debugging. In this paper, we introduce a DRL library \textit{FinRL} that facilitates beginners to expose themselves to quantitative finance and to develop their own stock trading strategies. Along with easily-reproducible tutorials, FinRL library allows users to streamline their own developments and to compare with existing schemes easily. Within FinRL, virtual environments are configured with stock market datasets, trading agents are trained with neural networks, and extensive backtesting is analyzed via trading performance. Moreover, it incorporates important trading constraints such as transaction cost, market liquidity and the investor's degree of risk-aversion. FinRL is featured with \textit{completeness, hands-on tutorial and reproducibility} that favors beginners: (i) at multiple levels of time granularity, FinRL simulates trading environments across various stock markets, including NASDAQ-100, DJIA, S\&P 500, HSI, SSE 50, and CSI 300; (ii) organized in a layered architecture with modular structure, FinRL provides fine-tuned state-of-the-art DRL algorithms (DQN, DDPG, PPO, SAC, A2C, TD3, etc.), commonly-used reward functions and standard evaluation baselines to alleviate the debugging workloads and promote the reproducibility, and (iii) being highly extendable, FinRL reserves a complete set of user-import interfaces.  Furthermore, we incorporated three application demonstrations, namely single stock trading, multiple stock trading, and portfolio allocation. The FinRL library will be available on Github at link https://github.com/AI4Finance-LLC/FinRL-Library.
  
\end{abstract}

\section{Introduction}

Deep reinforcement learning (DRL), which balances exploration (of uncharted territory) and exploitation (of current knowledge), has been recognized as an advantageous approach for automated stock trading. DRL framework is powerful in solving dynamic decision making problems by learning through interaction with an unknown environment, and thus providing two major advantages - portfolio scalability and market model independence \cite{buehler2019deep}. In quantitative finance, stock trading is essentially making dynamic decisions, namely \textit{to decide where to trade, at what price, and what quantity}, over a highly stochastic and complex stock market. As a result, DRL provides useful toolkits for stock trading \cite{hull2009options,xiong2018practical,zhang2020deep,yang2020,Dang2019ReinforcementLI,DRL_automate,li_a2c_2018}. Taking many complex financial factors into account, DRL trading agents build a multi-factor model and provide algorithmic trading strategies, which are difficult for human traders \cite{bekiros2010fuzzy,zhang2017online,kim2017intelligent,Jiang2017ADR}.

Preceding DRL, conventional reinforcement learning (RL) \cite{sutton2018reinforcement} has been applied to complex financial problems \cite{luenberger1997investment}, including option pricing, portfolio optimization and risk management. Moody and Saffell \cite{moody2001learning} utilized policy search and direct RL for stock trading. Deng \textit{et al.} \cite{Deng2017DeepDR} showed that applying deep neural networks profits more. There are industry practitioners who have explored trading strategies fueled by DRL, since deep neural networks are significantly powerful at approximating the expected return at a state with a certain action. With the development of more robust models and strategies, general machine learning approaches and DRL methods in specific are becoming more reliable. For example, DRL has been implemented on sentimental analysis on portfolio allocation \cite{li2019optimistic,Jiang2017ADR} and liquidation strategy analysis \cite{bao2019multi}, showing the potential of DRL on various financial tasks.

However, to implement a DRL or RL driven trading strategy is nowhere near as easy. The development and debugging processes are arduous and error-prone. Training environments, managing intermediate trading states, organizing training-related data and standardizing outputs for evaluation metrics - these steps are standard in implementation yet time-consuming especially for beginners. Therefore, we come up with a beginner-friendly library with fine-tuned standard DRL algorithms. It has been developed under three primary principles:
\begin{itemize}
  \item \textbf{Completeness}. Our library shall cover components of the DRL framework completely, which is a fundamental requirement;
  \item \textbf{Hands-on tutorials}. We aim for a library that is friendly to beginners. Tutorials with detailed walk-through will help users to explore the functionalities of our library;
  \item \textbf{Reproducibility}. Our library shall guarantee reproducibility to ensure the transparency and also provide users with confidence in what they have done.
\end{itemize}
%This spawns a unified and complete library where developers are able to efficiently explore ideas through high-level model operations, and customize their own environment modules when necessary.

In this paper, we present a three-layered \textit{FinRL} library that streamlines the development stock trading strategies. FinRL provides common building blocks that allow strategy builders to configure stock market datasets as virtual environments, to train deep neural networks as trading agents, to analyze trading performance via extensive backtesting, and to incorporate important market frictions. On the lowest level is environment, which simulates the financial market environment using actual historical data from six major indices with various environment attributes such as closing price, shares, trading volume, technical indicators etc. In the middle is the agent layer that provides fine-tuned standard DRL algorithms (DQN \cite{lillicrap2015continuous}\cite{mnih2015human}, DDPG \cite{lillicrap2015continuous}, Adaptive DDPG \cite{li2019optimistic}, Multi-Agent DDPG \cite{lowe2017multi}, PPO \cite{schulman2017proximal}, SAC \cite{haarnoja2018soft}, A2C \cite{mnih2016asynchronous} and TD3 \cite{Dankwa2019TwinDelayedDA}, etc.), commonly used reward functions and standard evaluation baselines to alleviate the debugging workloads and promote the reproducibility. The agent interacts with the environment through properly defined reward functions on the state space and action space. The top layer includes applications in automated stock trading, where we demonstrate three use cases, namely single stock trading, multiple stock trading and portfolio allocation. 

The contributions of this paper are summarized as follows:
\begin{itemize}
  \item FinRL is an open source library specifically designed and implemented for quantitative finance. Trading environments incorporating market frictions are used and provided.   
  %designed specifically for automated stock trading
  %\item Customization of trading time steps is feasible.
  \item Trading tasks accompanied by hands-on tutorials with built-in DRL agents are available in a beginner-friendly and reproducible fashion using Jupyter notebook. Customization of trading time steps is feasible.
  \item FinRL has good scalability, with a broad range of fine-tuned state-of-the-art DRL algorithms. Adjusting the implementations to the rapid changing stock market is well supported.
  \item Typical use cases are selected and used to establish a benchmark for the quantitative finance community. Standard backtesting and evaluation metrics are also provided for easy and effective performance evaluation. 
\end{itemize}

The remainder of this paper is organized as follows. Section 2 reviews related works. Section 3 presents FinRL Library. Section 4 provides evaluation support for analyzing stock trading performance. We conclude our work in Section 5.

\section{Related Works}

We review related works on relevant open source libraries and existing applications of DRL in finance.  

\subsection{State-of-the-Art Algorithms}

Recent works can be categorized into three approaches: value based algorithm, policy based algorithm, and actor-critic based algorithm. FinRL has consolidated and elaborated upon those algorithms to build financial DRL models. There are a number of machine learning libraries that share similar features as our FinRL library. \
%In Table 1 we compare with them from different perspectives.

%\begin{table*}[htb]
%\caption{Comparison of RL Libraries}
%\begin{tabular}{|l|l|l|l|l|}
%\hline
%\textbf{Libraries}       & \textbf{SOTA RL}        & %\textbf{Doc.}           & \textbf{Pip Installer}        & \textbf{Finance Customization}       \\ \hline
%\textbf{FinRL}       & Yes     & Yes     & Yes     & Yes     %    \\ \hline
%\textbf{OpenAI Gym}       & Yes     & No     & Yes     & No  %       \\ \hline
%\textbf{Google Dopamine}         & Yes     & Yes     & Yes   %  & No         \\ \hline
%\textbf{RLlib}          & Yes     & Yes     & Yes     & No   %      \\ \hline
%\textbf{RLlab}             & Yes     & Yes     & Yes     & No %        \\ \hline
%\textbf{Personae}         & Yes     & No     & Yes     & No  %       \\ \hline
%\textbf{Tensorforce}        & Yes     & Yes     & Yes     & No         \\ \hline
%\textbf{Horizon}        & Yes     & No     & No     & No         \\ \hline
%\end{tabular}
%\end{table*}

\begin{itemize}
\item \textbf{OpenAI Gym} \cite{brockman2016openai} is a popular open source library that provides a standardized set of task environments. OpenAI Baselines \cite{baselines} implements high quality deep reinforcement learning algorithms using gym environments. Stable Baselines \cite{stable-baselines} is a fork of OpenAI Baselines with code cleanup and user-friendly examples.

\item \textbf{Google Dopamine} \cite{castro18dopamine} is a research framework for fast prototyping of reinforcement learning algorithms. It features plugability and reusability. 

\item \textbf{RLlib} \cite{liang2018rllib} provides high scalability with reinforcement learning algorithms. It has modular framework and is very well maintained.

%\item \textbf{TensorLayer} \cite{dong2017tensorlayer} is a deep learning library with RL support, designed for researchers and engineers to customize neural networks for real-world applications. TensorLayer is a wrapper of TensorFlow and uses the OpenAI Gym module as environment, not user-friendly for financial tasks.

%\item \textbf{RLlab} \cite{duan2016benchmarking} is a reinforcement learning framework for developing continuous control tasks. 

%\item \textbf{Personae} \cite{Personae} includes DRL algorithms and applies them to financial tasks. It includes a basic environment simulating financial market supporting stock and future. However, it lacks market frictions, nor does it allow customization. 

%\item \textbf{Tensorforce} \cite{lift-tensorforce} Tensorforce is an extension to Tensorflow and gains all benefits of using Tensorflow. However, the code is complicated which makes it not easy for beginners to get hands on.

\item \textbf{Horizon} \cite{gauci2018horizon}  is a DL-focused framework dominated by PyTorch, whose main use case is to train RL models in the batch setting.
\end{itemize}

%%%%%%%%%%%%%%%%%%%%%%%%%%%%%%%%%%%%%%%%%%%%%%%%%%%
\subsection{DRL in Finance}

Recent works show that DRL has many applications in quantitative finance \cite{RL_survey}.

Stock trading is usually considered as one of the most challenging applications due to its noisy and volatile features. Many researchers have explored various approaches using DRL \cite{nechchi2016reinforcement, Nan2020SentimentAK, Dang2019ReinforcementLI, Corazza2014QLearningBasedFT, zhang2020deep, Ganesh2018DeepRL}. Volatility scaling can be incorporated with DRL to trade futures contracts \cite{zhang2020deep}. By adding a market volatility term to reward functions, we can scale up the trade shares with low volatility, and vice versa. News headline sentiments and knowledge graphs can also be combined with the time series stock data to train an optimal policy using DRL \cite{Nan2020SentimentAK}. High frequency trading with DRL is also a hot topic \cite{Ganesh2018DeepRL}. Deep Hedging \cite{buehler2019deep, Cao2019DeepHO} represents hedging strategies with neural networks learned by modern DRL policy search. This application has shown two key advantages of the DRL approach in quantitative finance, which are scalability and model independent. It uses DRL to manage the risk of liquid derivatives, which indicates further extension of our library into other asset classes and topics. 

%Cryptocurrency is considered as a significant and rising element in today's digital financial market. Crypto markets led by Bitcoin (BTC)  \cite{Sapuric2014BitcoinIV} are considered as more volatile than stock market. We can also apply DRL to those markets performing automated trading \cite{Lucarelli2019ADR}, portfolio allocation \cite{Jiang2017CryptocurrencyPM}, and market making \cite{Sadighian2019DeepRL}.

\section{The Proposed FinRL Library}

FinRL library consists of three layers: environments, agents and applications. We first describe the overall architecture, and then present each layer.
%the simulated trading environment layer and DRL trading agent layer, respectively.

\subsection{Architecture of the FinRL Library}

\begin{figure}[t]
\centering
\includegraphics[scale=1,width=1\textwidth]{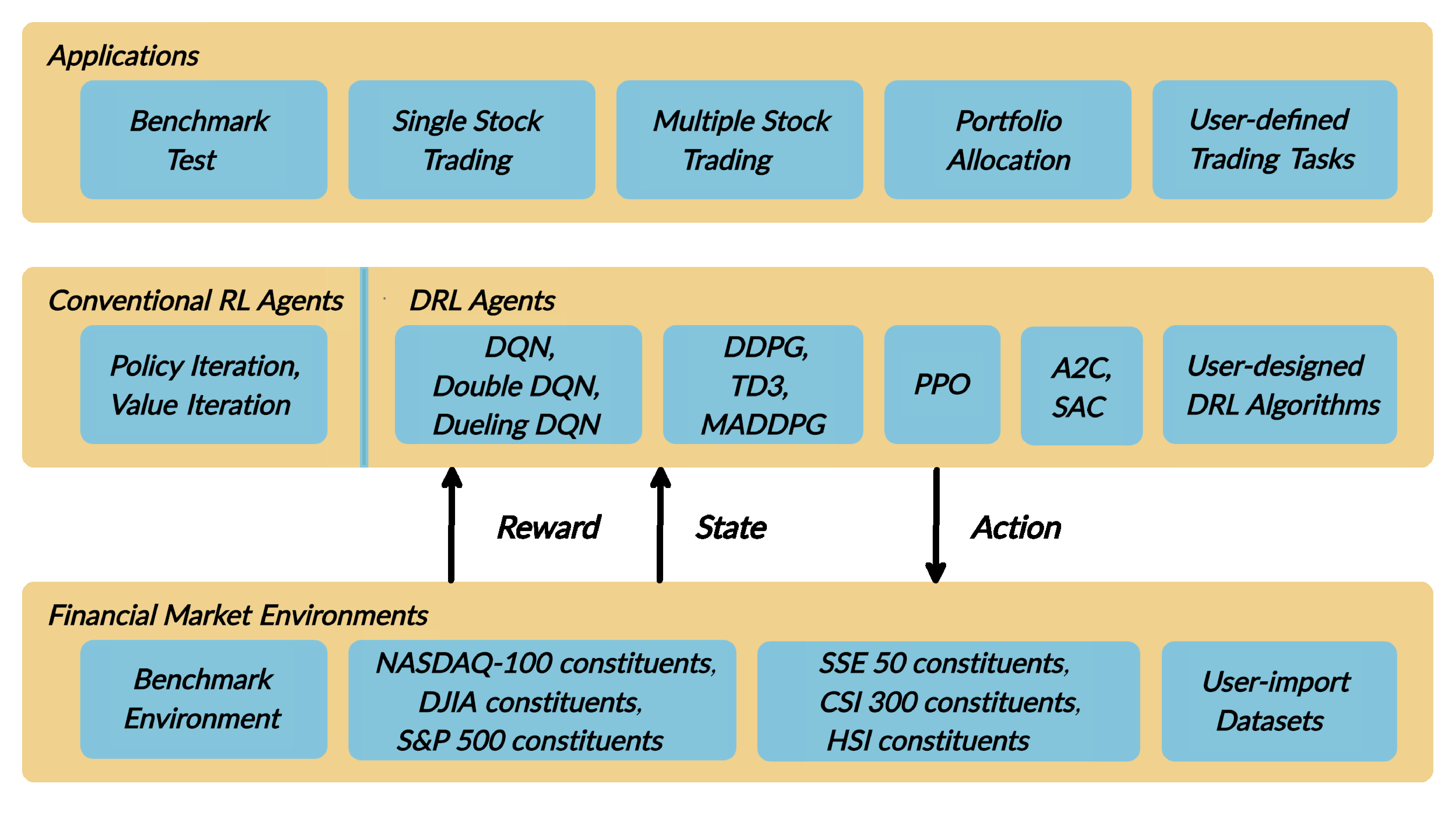}
\caption{An overview of our FinRL library. It consists of three layers: application layer, DRL agent layer, and the finance market environment layer.}
\label{overview}
\end{figure}

The architecture of the FinRL library is shown in Fig. \ref{overview}, and its features are summarized as follows:
\begin{itemize}
    \item \textbf{Three-layer architecture}: The three layers of FinRL library are stock market environment, DRL trading agent, and stock trading applications. The agent layer interacts with the environment layer in an exploration-exploitation manner, whether to repeat prior working-well decisions or to make new actions hoping to get greater rewards. The lower layer provides APIs for the upper layer, making the lower layer transparent to the upper layer.
    \item \textbf{Modularity}: Each layer includes several modules and each module defines a separate function. One can select certain modules from any layer to implement his/her stock trading task. Furthermore, updating existing modules is possible.
    \item \textbf{Simplicity, Applicability and Extendibility}: Specifically designed for automated stock trading, FinRL presents DRL algorithms as modules. In this way, FinRL is made accessible yet not demanding. FinRL provides three trading tasks as use cases that can be easily reproduced. Each layer includes reserved interfaces that allow users to develop new modules. 
    \item \textbf{Better Market Environment Modeling}: We build a trading simulator that replicates live stock market and provides backtesting support that incorporates important market frictions such as transaction cost, market liquidity and the investor's degree of risk-aversion. All of those are crucial among key determinants of net returns. 
\end{itemize}

%Each layer of FinRL library is described as follows.

%On the environment layer, six frequently-referred-to index, NASDAQ-100, DJIA, S\&P 500, SSE 50, CSI 300, HSI are included. 

%On the agent layer, FinRL provides both RL and DRL fine-tuned algorithms, which can be directly used for and tested in user's stock trading tasks. Conventional learning algorithms include value iteration and policy iteration, which provide validations on small data sets for developing DRL algorithms. DRL agents include DQN \cite{mnih2015human}, DDPG \cite{lillicrap2015continuous}, Adaptive DDPG \cite{li2019optimistic}, Multi-Agent DDPG \cite{lowe2017multi}, PPO \cite{schulman2017proximal}, SAC \cite{haarnoja2018soft}, A2C \cite{mnih2016asynchronous} and TD3 \cite{fujimoto2018addressing} algorithms. 

%On the application layer, FinRL provides application demonstrations and benchmark tests. Benchmark tests include both conventional RL \& DRL algorithms that are trained with tens of stocks. The benchmark tests serve as preliminary verification for users when they develop their own strategies. The application demonstrations include single stock trading using DQN and DDPG, portfolio allocation using Adaptive DDPG, and liquidation strategy analysis using MADDPG.

%\begin{figure}[t]
%\centering
%\includegraphics[height=3.5cm]{image/state521.png}
%\caption{State transition diagram.}
%\label{fig2}
%\end{figure}

%frame530

\subsection{Environment: Time-driven Trading Simulator}

Considering the stochastic and interactive nature of the automated stock trading tasks, a financial task is modeled as a Markov Decision Process (MDP) problem. The training process involves observing stock price change, taking an action and reward's calculation to have the agent adjusting its strategy accordingly. By interacting with the environment, the trading agent will derive a trading strategy with the maximized rewards as time proceeds.

%introduction
Our trading environments, based on OpenAI Gym framework, simulate live stock markets with real market data according to the principle of time-driven simulation \cite{brockman2016openai}. FinRL library strives to provide trading environments constructed by six datasets across five stock exchanges. 
%The state space representing the environment is used to pick the action in the action state and the reward as incentive. Every trading day, price of stocks in the state space changes and the agent takes trading actions on this basis.  

%It serves with supervision under real market and is more realistic and robust compared to artificial environment for DRL.
%why real market
%data set 

\subsubsection{State Space, Action Space, and Reward Function} We give definitions of the state space, action space and reward function.

\textbf{State space $\mathcal{S}$}.
The state space describes the observations that the agent receives from the environment. Just as a human trader needs to analyze various information before executing a trade, so our trading agent observes many different features to better learn in an interactive environment. We provide various features for users:
\begin{itemize}
\item Balance ${b}_{t}\in \mathbb{R}_+$: the amount of money left in the account at the current time step $t$.
\item Shares own $\bm{h}_{t}\in \mathbb{Z}_+^{n}$: current shares for each stock, $n$ represents the number of stocks.
\item Closing price $\bm{p}_{t}\in \mathbb{R}_+^{n}$: one of the most commonly used feature.
\item Opening/high/low prices $\bm{o}_{t}, \bm{h}_{t}, \bm{l}_{t}\in \mathbb{R}_+^{n}$: used to track stock price changes.
\item Trading volume $\bm{v}_{t}\in \mathbb{R}_+^{n}$: total quantity of shares traded during a trading slot.
\item Technical indicators: Moving Average Convergence Divergence (MACD) $\bm{M}_t \in \mathbb{R}^{n}$ and Relative Strength Index (RSI) $\bm{R}_t \in \mathbb{R}_+^{n}$, etc.
\item Multiple-level of granularity: we allow data frequency of the above features to be daily, hourly or on a minute basis.
\end{itemize}

\textbf{Action space $\mathcal{A}$}. 
The action space describes the allowed actions that the agent interacts with the environment. Normally, $a \in \mathcal{A}$ includes three actions: $a \in \{-1, 0, 1\}$,  where $-1, 0, 1$ represent selling, holding, and buying one stock. Also, an action can be carried upon multiple shares. We use an action space $\{-k,...,-1, 0, 1, ..., k\}$, where $k$ denotes the number of shares. For example, "Buy 10 shares of AAPL" or "Sell 10 shares of AAPL" are $10$ or $-10$, respectively.
%A continuous action space needs to be normalized to $[-1, 1]$, since the policy is defined on a Gaussian distribution, which needs to be normalized and symmetric.

%The reward function $r(s,a,s')$

\textbf{Reward function $r(s, a, s')$} is the incentive mechanism for an agent to learn a better action. There are many forms of reward functions. We provide commonly used ones \cite{RL_survey} as follows:
\begin{itemize}
\item The change of the portfolio value when action $a$ is taken at state $s$ and arriving at new state $s'$ \cite{Deng2017DeepDR, xiong2018practical,Dang2019ReinforcementLI,Nan2020SentimentAK, yang2020}, i.e., $r(s,a,s') =  v' - v$,
where $v'$ and $v$ represent the portfolio values at state $s'$ and $s$, respectively.
\item The portfolio log return when action $a$ is taken at state $s$ and arriving at new state $s'$ \cite{huang2018financial}, i.e., $r(s,a,s') =  \log(\frac{v'}{v})$.
\item The Sharpe ratio for periods $t=\{1,...,T\}$ \cite{jin2016portfolio, Moody1998PerformanceFA}, i.e., $S_T = \frac{\text{mean}(R_t)}{\text{std}(R_t)}$, where $ R_t =  v_t - v_{t-1}$. 

\item FinRL also supports user defined reward functions to include risk factor or transaction cost term such as in \cite{Deng2017DeepDR,zhang2020deep,buehler2019deep}
\end{itemize} 

%\textbf{Action-value function $Q_{\pi}(s, a)$}.
%Action-value function $Q_{\pi}(s, a)$ means the expected value of taking action $a$ in state $s$ under policy $\pi$.
\subsubsection{Standard and User Import Datasets}

The application of DRL in finance is different from that in other fields, such as playing chess and card games \cite{silver2016mastering,Zha2019ExperienceRO}; the latter inherently have clearly defined rules for environments. 
Various finance markets require different DRL algorithms to get the most appropriate automated trading agent. 
Realizing that setting up training environment is a time-consuming and laborious work, FinRL provides six environments based on representative listings, including NASDAQ-100, DJIA, S\&P 500, SSE 50, CSI 300, and HSI, plus one user-defined environment. With those efforts, this library frees users from tedious and time-consuming data pre-processing workload.

We are well aware that users may want to train trading agents on their own data sets. FinRL library provides convenient support to user imported data to adjust the granularity of time steps. We specify the format of the data for each of the use cases. Users only need to pre-process their data sets according to our data format instructions. 

\subsection{Deep Reinforcement Learning Agents}

\begin{figure*}[t]
\centering
\includegraphics[width=1.15\textwidth]{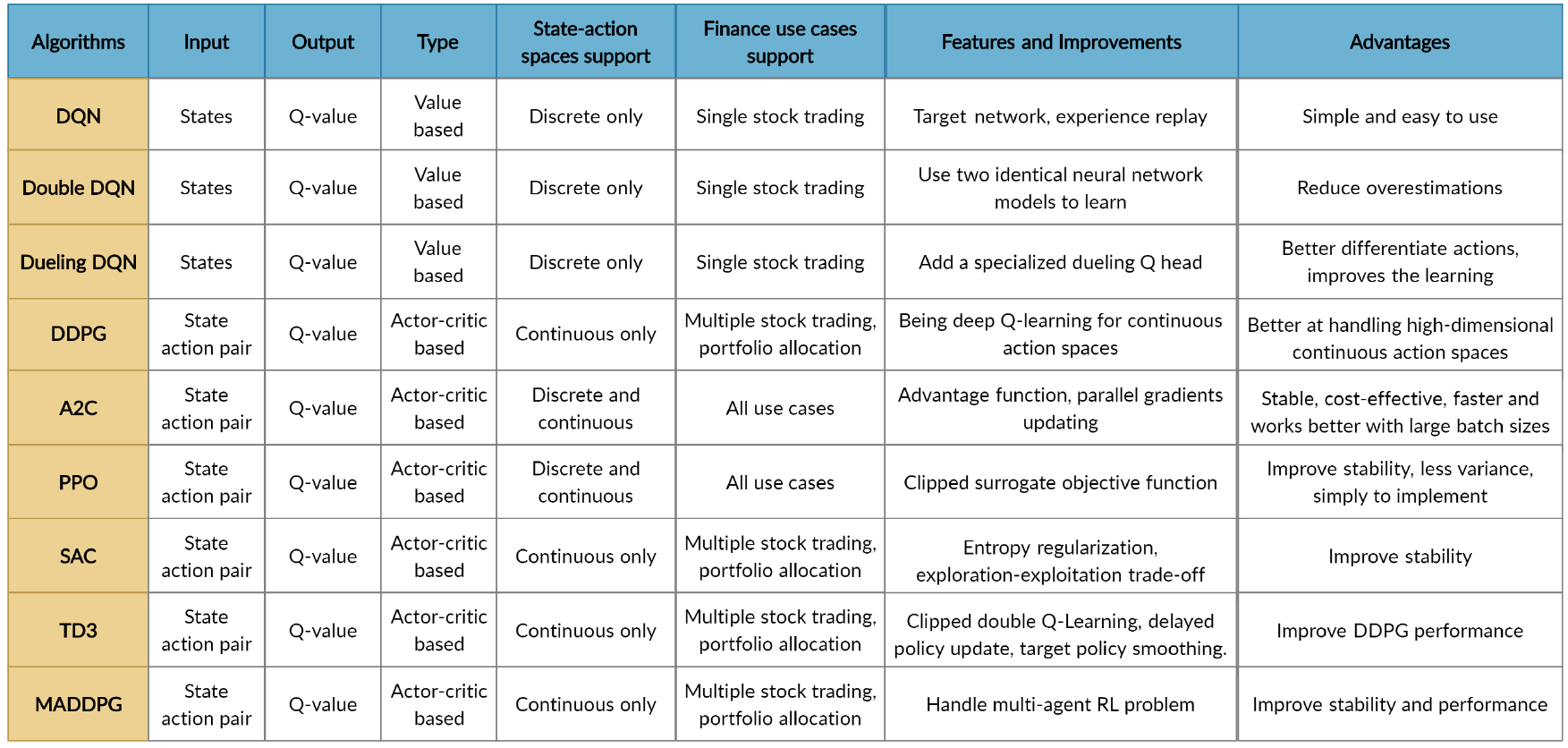}
\caption{Comparison of DRL algorithms.}
\label{compare}
\end{figure*}

FinRL library includes fine-tuned standard DRL algorithms, namely, DQN \cite{lillicrap2015continuous}\cite{mnih2015human}, DDPG \cite{lillicrap2015continuous}, Multi-Agent DDPG \cite{lowe2017multi}, PPO \cite{schulman2017proximal}, SAC \cite{haarnoja2018soft}, A2C \cite{mnih2016asynchronous} and TD3 \cite{Dankwa2019TwinDelayedDA}. We also allow users to design their own DRL algorithms by adapting these DRL algorithms, e.g., Adaptive DDPG \cite{li2019optimistic}, or employing ensemble methods \cite{yang2020}. The comparison of DRL algorithms is shown in Fig. \ref{compare}

The implementation of the DRL algorithms are based on OpenAI Baselines \cite{baselines} and Stable Baselines \cite{stable-baselines}.

\section{Evaluation of Trading Performance}

Standard metrics and baseline trading strategies are provided to support trading performance analysis. FinRL library follows a training-validation-testing flow to design a trading strategy. 
%FinRL also considers many trading constraints to perform backtesting tests.

\subsection{Standard Performance Metrics}
FinRL provides five evaluation metrics to help users evaluate the stock trading performance directly, which are final portfolio value, annualized return, annualized standard deviation, maximum drawdown ratio, and Sharpe ratio.

%\textbf{Final portfolio value}: portfolio value at the end of the trading stage.
%
%\textbf{Annualized return}: geometric average amount of money earned by the agent each %year over the time period. 
%
%\textbf{Annualized standard deviation}: annualized standard deviation of the portfolio %return $R_{p}$.
%
%\textbf{Maximum drawdown ratio}: the maximum percentage loss during the trading eriod.
%% FinRL pushes back at any historical point in the selected cycle, then find the nadir %of total assets. 
%
%\textbf{Sharpe ratio:} the average return earned in excess of the risk-free rate per %unit of volatility or total risk:
%\begin{equation}
%\text{Sharpe Ratio} =\frac{\mathbb{E}\left(R_{p}\right)-R_{f}}{\sigma_{p}},
%\end{equation}
%where $\mathbb{E}\left(R_{p}\right)$ is expected return rate, $R_{f}$ is risk-free %rate, and $\sigma_{p}$ is the standard deviation of $R_{p}$. 
 
\subsection{Baseline Trading Strategies}
Baseline trading strategies should be well-chosen and follow industrial standards. The strategies will be universal to measure, standard to compare with, and easy to implement. In FinRL library, traditional trading strategies serve as the baseline for comparing with DRL strategies. Investors usually have two objectives for their decisions: the highest possible profits and the lowest possible risks of uncertainty \cite{sharpe1970portfolio}. FinRL uses five conventional strategies, namely passive buy-and-hold trading strategy \cite{Malkiel2003PassiveIS}, mean-variance strategy \cite{meanvariance2012}, and min-variance strategy \cite{meanvariance2012}, momentum trading strategy \cite{Foltice2015ProfitableMT}, and equal-weighted strategy to address these two mutually limiting objectives and the industrial standards.

%\textbf{Passive Trading Strategy} \cite{Malkiel2003PassiveIS}: is a popular, very easy to implement, and lucrative strategy among investors. With minimal trading activities, investors simply buy and hold index ETFs \cite{Tarassov2016ExchangeTF} to replicate a broad market index or indices. Popular index ETFs are: SPDR S\&P 500 ETF Trust (SPY) and Vanguard 500 Index Fund (VOO) for S\&P 500 Index, SPDR Dow Jones Industrial Average ETF Trust (DIA) for Dow Jones Industrial Average, Invesco QQQ Trust Series 1 (QQQ) for Nasdaq 100 index, etc.

%\textbf{Mean-Variance Strategy} \cite{meanvariance2012}: to achieve an optimal balance between the risks and profits, or the highest Sharpe ratio. The objective of the mean-variance strategy is to minimize risk for a given level of expected return:
%\begin{equation}
%\begin{split}
%&\underset{{w_i}}{\min}~\text{var}(r_p) \\
%&\text{subject~to}~\mathbb{E}(r_p) = \mu^*,~~ \sum w_i = 1,
%
%\end{split}
%\end{equation} where $r_p$ is the portfolio return, $w_i$ is the portfolio weights, and %$\mu^*$ is the expected or target return.

%\textbf{Min-Variance Strategy} \cite{meanvariance2012}: to find the portfolio with the smallest variance. It indicates a well-diversified portfolio that consists of individually risky assets, which are hedged when traded together, resulting in the lowest possible risk for the rate of expected return. The objective function of min-variance model is similar to mean-variance model without setting the target return.

%%%%%%%%%%%%%%%%%%%%%%%%%%%%%%%%%%%%%%
\subsection{Training-Validation-Testing Flow}

\begin{figure}[t]
\centering
\includegraphics[scale=0.46]{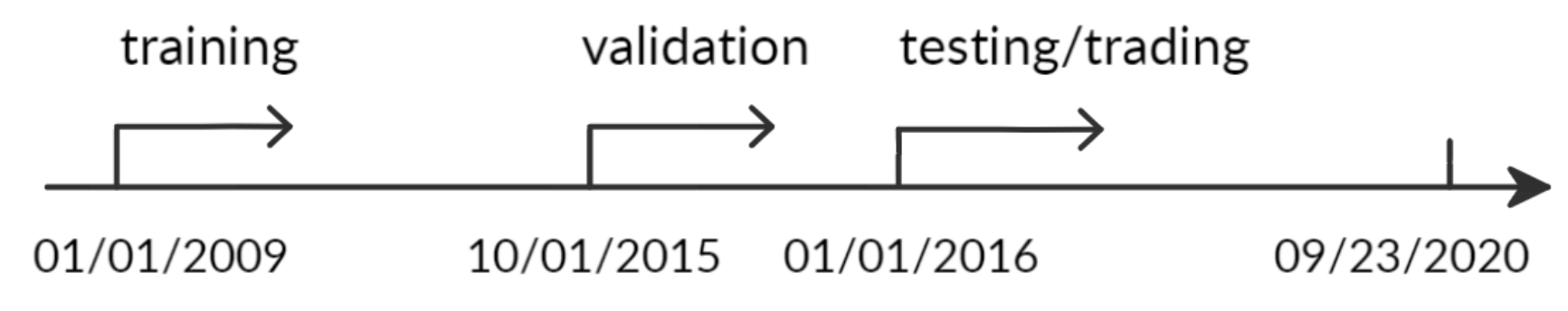}
\caption{Data splitting.}
\label{data}
\end{figure}

%\begin{figure}[t]
%\centering
%\includegraphics[height=4.5cm]{image/MADDPG515.png}
%\caption{The liquidation process with independent, competitive and cooperative agents using the FinRL library.}
%\label{fig5}
%\vspace{-0.17in}
%\end{figure}

%\begin{figure*}[t]
%\centering
%\includegraphics[height=2.0cm]{image/timeline507.png}
%\caption{Data splitting.}
%\label{fig4}
%\end{figure*}

%\yanglet{To Bruce: need to give reasons about why this flow is necessary for financial tasks?}
With our use cases as instances, the stock market data are divided into three phases in Fig. \ref{data}. Training dataset is the sample of data to fit the DRL model. The model sees and learns from the training dataset. Validation dataset is used for parameter tuning and to avoid overfitting. Testing (trading) dataset is the sample of data to provide an unbiased evaluation of a fine-tuned model. Rolling window is usually associated with the training-validation-testing flow in stock trading because investors and portfolio managers may need to rebalance the portfolio and retrain the model periodically. FinRL provides flexible rolling window selection such as on a daily basis, monthly, quarterly, yearly or by user specified.

%Specifically, with our use cases as instances, the stock market data are divided into three phases in Figure \ref{data}.
%According to the standard workflow in quantitative finance, 

%\textbf{Training phase}: we use data from 01/01/2009 to 09/30/2015 to train the automated trading agents.

%\textbf{Validation phase}: we use data from 10/01/2015 to 12/31/2015 to validate the results and adjust key parameters, such as time slice's size and learning rate.

%\textbf{Testing/Trading phase}: we use data from 01/01/2016 to 05/08/2020 (unseen data) to evaluate the performance of the applications.

%In performance evaluation of RL algorithms, the portfolio value and stock index yield are compared in the form of curve graphs in fig.3), users can see the performance of the algorithm on the DJIA constituents' data. Then specific indicators are shown in Table 1, showing the risks of trading using algorithms.

%From the results of RL, we can conclude that DRL method have better performance on DJIA than traditional investment methods and RL methods. There are also differences among different DRL methods. The evaluation model of FinRL is multi-angled and effective.

%%%%%%%%%%%%%%%%%%%%%%%%%%%%%%%%%%%%%%
\subsection{Backtesting with Trading Constraints}

In order to better simulate practical trading, we incorporate trading constraints, risk-aversion and automated backtesting tools.

\textbf{Automated Backtesting}.
Backtesting plays a key role in performance evaluation. Automated backtesting tool is preferable because it reduces the human error. In FinRL library, we use the Quantopian pyfolio package \cite{pyfolio} to backtest our trading strategies. This package is easy to use and consists of various individual plots that provide a comprehensive image of the performance of a trading strategy. 

\textbf{Incorporating Trading Constraints}.
Transaction costs incur when executing a trade. There are many types of transaction costs, such as broker commissions and SEC fee. We allow users to treat transaction costs as a parameter in our environments:
\begin{itemize}
    \item Flat fee: a fixed dollar amount per trade regardless of how many shares traded.
    \item Per share percentage: a per share rate for every share traded, for example, 1/1000 or 2/1000 are the most commonly used transaction cost rate for each trade.
\end{itemize}
Moreover, we need to consider market liquidity for stock trading, such as bid-ask spread. Bid-ask spread is the difference between the prices quoted for an immediate selling action and an immediate buying action for stocks. In our environments, we can add the bid-ask spread as a parameter to the stock closing price to simulate real world trading experience. 
%Lastly, in order to address the risk-aversion of various traders, one may add the standard deviation of the portfolio returns into our reward function or simply use the risk-adjusted Sharpe ratio as the reward function.

\begin{figure*}[t]
\centering
\includegraphics[width=1\textwidth]{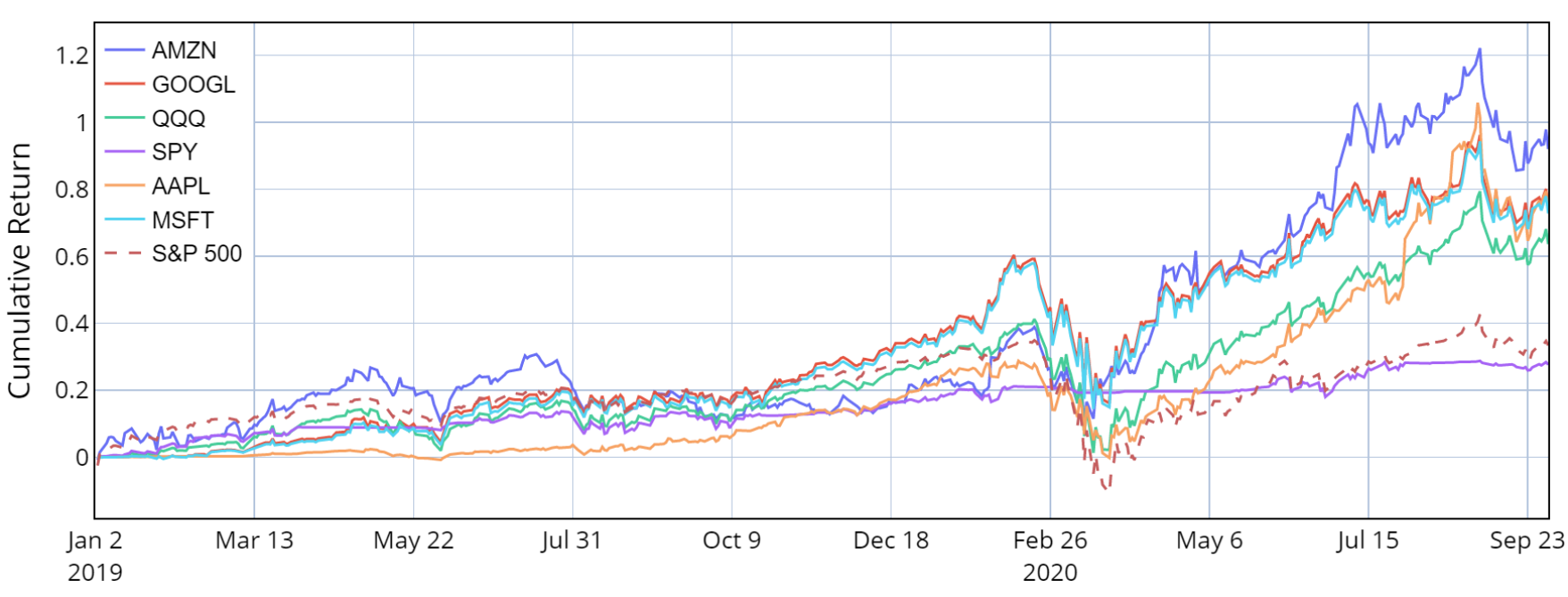}
\caption{Performance of single stock trading using PPO in the FinRL library.}
\label{single_stock}
\end{figure*}

\begin{figure*}[t]
\centering
\includegraphics[width=1\textwidth]{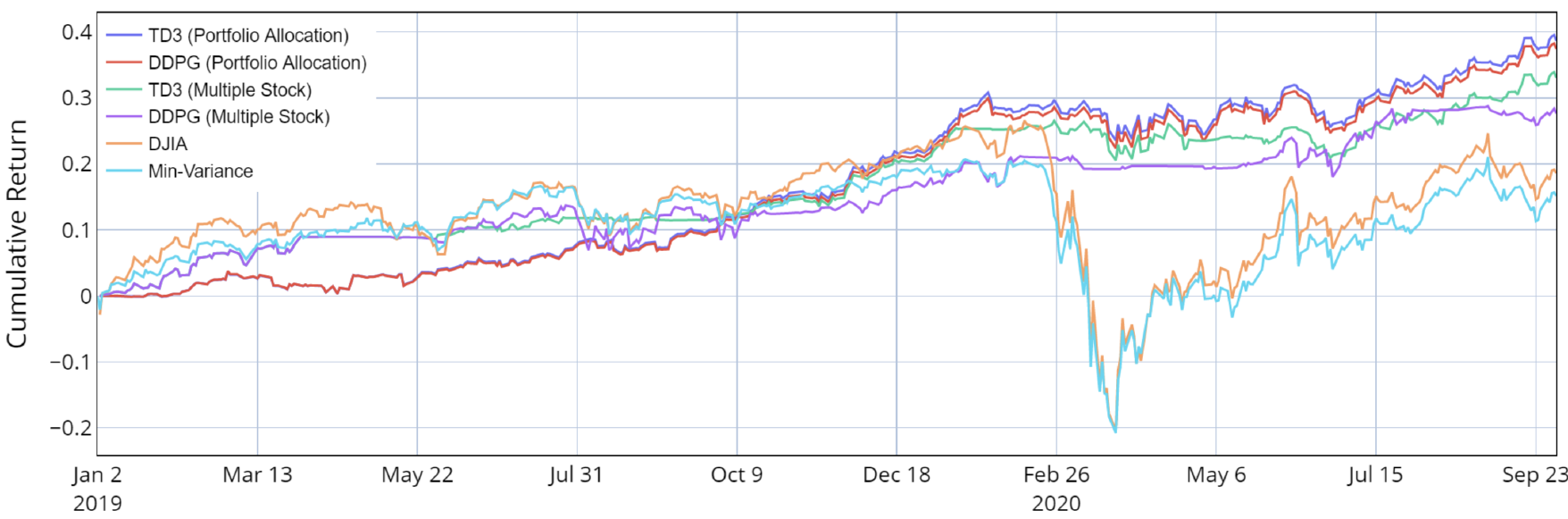}
\caption{Performance of multiple stock trading and portfolio allocation using the FinRL library.}
\label{fig3}
\end{figure*}

\begin{table*}[t]
		\centering
		\resizebox{\textwidth}{!}{
		\begin{tabular}{|l|c|c|c|c|c|c|c|c|c|c|c|}\hline
			2019/01/01-2020/09/23 & SPY  & QQQ&GOOGL & AMZN & AAPL  &  MSFT  &  S\&P 500 \\
			\hline
			Initial value   
			&100,000 &100,000&100,000&100,000&100,000&100,000&100,000\\
			Final value     
			&127,044&163,647&174,825&192,031&173,063&172,797&133,402\\
			Annualized return 
			&14.89\%&32.33\% &37.40\%& 44.94\% &36.88\%&36.49\% &17.81\% \\
			Annualized Std &9.63\%&27.51\%&33.41\%&29.62\%&25.84\%&33.41\%&27.00\%\\
			Sharpe ratio      
			& 1.49 & 1.16& 1.12& 1.40& 1.35 & 1.10 &  0.74   \\
			Max drawdown &20.93\%&28.26\%&27.76\%&21.13\%&22.47\%&28.11\%&33.92\%\\
			\hline
		\end{tabular}}
		\caption{Performance of single stock trading using PPO in the FinRL library. The Sharpe ratio of all the ETFs and stocks outperform the market, namely the S\&P 500 index.}
		\label{single_stock_performance}
	\end{table*}

\begin{table*}[t]
		\centering
		\resizebox{\textwidth}{!}{
		\begin{tabular}{|l|c|c|c|c|c|}\hline
			2019/01/01-2020/09/23 & TD3 & DDPG&  Min-Var. &  DJIA \\
			\hline
			Initial value   
			&1,000,000 &1,000,000&1,000,000&1,000,000\\
			Final value     
			&\textcolor{orange}{1,403,337};~~\textcolor{cyan}{1,381,120}& \textcolor{orange}{1,396,607};~~\textcolor{cyan}{1,281,120}&1,171,120&1,185,260\\
			Annualized return 
			&\textcolor{orange}{21.40\%}; ~~\textcolor{cyan}{17.61\%}&
			\textcolor{orange}{20.34\%}; ~~\textcolor{cyan}{15.81\%}&8.38\%& 10.61\%\\
			Annualized Std 
			&\textcolor{orange}{14.60\%}; ~~\textcolor{cyan}{17.01\%}&
			\textcolor{orange}{15.89\%};~~\textcolor{cyan}{16.60\%}
			&26.21\%&28.63\%\\
			Sharpe ratio      
			& \textcolor{orange}{1.38}; ~~\textcolor{cyan}{1.03}
			&\textcolor{orange}{1.28}; ~~\textcolor{cyan}{0.98} & 0.44 &0.48  \\
			Max drawdown 
			&\textcolor{orange}{11.52\%}~~\textcolor{cyan}{12.78\%}&
			\textcolor{orange}{13.72\%}; ~~\textcolor{cyan}{13.68\%}&34.34\%&37.01\%\\
			\hline
		\end{tabular}}
		\caption{Performance of \textcolor{orange}{multiple stock trading} and \textcolor{cyan}{portfolio allocation} over the DJIA constituents stocks using the FinRL library. The Sharpe ratios of TD3 and DDPG excceed the DJIA index, and the traditional min-variance portfolio allocation strategy.}
		\label{tab:Performance evaluation}
		\vspace{-0.15in}
	\end{table*}
	
\textbf{Risk-aversion}.
Risk-aversion reflects whether an investor will choose to preserve the capital. It also influences one's trading strategy when facing different market volatility level.

To control the risk in a worst-case scenario, such as financial crisis of 2007–2008, FinRL employs the financial turbulence index $\textit{turbulence}_t$ that measures extreme asset price fluctuation \cite{turbulence}:
\begin{equation}  
\textit{turbulence}_t = \left(\bm{y_t} - \bm{\mu}\right)\bm{\Sigma^{-1}}(\bm{y_t}-\bm{\mu})' \in \mathbb{R},
\label{turb}
\end{equation} 
where $\bm{y_t} \in \mathbb{R}^n$ denotes the stock returns for current period t, $\bm{\mu} \in \mathbb{R}^n$ denotes the average of historical returns, and $\bm{\Sigma} \in \mathbb{R}^{n \times n}$ denotes the covariance of historical returns. It is used as a parameter that controls buying or selling action, for example if the turbulence index reaches a pre-defined threshold, the agent will halt buying action and starts selling the holding shares gradually.

\subsection{Demonstration of Three Use Cases}

We demonstrate with three use cases: single stock trading \cite{Dang2019ReinforcementLI,DRL_automate,li_a2c_2018,zhang2020deep}, multiple stock trading \
\cite{xiong2018practical,yang2020}, and portfolio allocation \cite{Jiang2017ADR,li2019optimistic}. FinRL library provides practical and reproducible solutions for each use case, with online walk-through tutorial using Jupyter notebook (e.g., the configurations of the running environment and commands). We select three use cases and reproduce the results using FinRL to establish a benchmark for the quantitative finance community.

Fig. \ref{single_stock} and Table \ref{single_stock_performance} demonstrate the performance evaluation of single stock trading. We pick large-cap ETFs such as SPDR S\&P 500 ETF Trust (SPY) and Invesco QQQ Trust Series 1 (QQQ), and stocks such as Google (GOOGL), Amazon (AMZN), Apple (AAPL), and Microsoft (MSFT). We use PPO algorithm in FinRL and train a trading agent.  The maximum drawdown in Table \ref{single_stock_performance} is large due to Covid-19 market crash.

Fig. \ref{fig3} and Table \ref{tab:Performance evaluation} show the performance and multiple stock trading and portfolio allocation over the Dow Jones 30 constitutes. We use DDPG and TD3 to trade multiple stocks, and allocate portfolio.

\section{Conclusions}

In this paper, we have presented FinRL library that is a DRL library designed specifically for automated stock trading with an effort for educational and demonstrative purpose.  FinRL is characterized by its extendability, more-than-basic market environment and extensive performance evaluation tools also for quantitative investors and strategy builders. Customization is easily accessible on all layers, from market simulator, trading agents' learning algorithms up towards profitable strategies. In a trading strategy design, FinRL follows a training-validation-testing flow and provides automated backtesting as well as benchmark tests. As a walk-through tutorial in Jupyter notebook format, we demonstrate easily reproducible profitable strategies under different scenarios using FinRL: (i) single stock trading; (ii) multiple stock trading; (iii) incorporating the mechanism of stock information penetration. With FinRL Library, implementation of powerful DRL driven trading strategies is made an accessible, efficient and delightful experience.

\newpage

\medskip
\small
\bibliographystyle{plain}
\bibliography{ref}

\end{document}